\documentclass[doublecol]{epl2} 

\usepackage{amsmath}
\usepackage{cleveref}

\crefname{table}{}{}
\Crefname{table}{}{}
\crefname{}{}{}
\Crefname{}{}{}
\crefname{equation}{}{}
\Crefname{Equation}{}{}
\newcommand{\mbf}{\boldsymbol}

\newcommand{\kper}{k_{\perp}}
\newcommand{\kpar}{k_{\parallel}}
\newcommand{\bk}{\mbf{k}}

\newcommand{\bu}{\mbf{u}}
\newcommand{\bU}{\mbf{U}}

\newcommand{\TKQ}{\mathcal{T}}
\newcommand{\bnabla}{\mbf{\nabla}}
\newcommand{\ZK}{Zakharov--Kolmogorov }

\newcommand{\Eddd}{\mathcal{E}_{\mathrm{3d}}}
\newcommand{\Edd}{\mathcal{E}_{\mathrm{2d}}}
\title{Evidence of the \ZK spectrum in numerical simulations of inertial wave turbulence}
\shorttitle{Evidence of the \ZK spectrum of inertial wave turbulence} %Insert here a short version of the title if it exceeds 70 characters

\author{T. Le Reun \inst{1,2} \and B. Favier \inst{1} \and M. Le Bars\inst{1}}
\shortauthor{T. Le Reun \etal}

\institute{                    
  \inst{1}  Aix Marseille Univ, CNRS, Centrale Marseille, IRPHE UMR 7342, Marseille, France\\
  \inst{2} DAMTP, University of Cambridge, Wilberforce Road, Cambridge CB3 0WA, UK
}
\abstract{
Rotating turbulence is commonly known for being dominated by geostrophic vortices that are invariant along the rotation axis and undergo inverse cascade.
Yet, it has recently been shown to sustain fully three-dimensional states with a downscale energy cascade.
In this letter, we investigate the statistical properties of three-dimensional rotating turbulence by the means of direct numerical simulations in a triply periodic box where geostrophic vortices are specifically damped. 
The resulting turbulent flow is an inertial wave turbulence that verifies the \ZK spectrum derived analytically by Galtier (Phys. Rev. E, 68, 2003), thus offering numerical proof of the relevance of wave turbulence theory for three-dimensional, anisotropic waves.
\revision{Lastly, we show that the same forcing leads to either geostrophic or wave turbulence depending on the initial condition.}
Our results thus bring further evidence for bi-stability in rotating turbulent flows. 
}

\begin{document}
\renewcommand{\ref}{\cref}

\maketitle

\section{Introduction} 

% Plan
% 1. Importance de la turbulence en rotation
% 2. 
%
%
%

Rotating turbulent flows are ubiquitous in geo- and astrophysical systems such as stellar interiors, planetary cores, oceans and atmosphere. 
One of its most well known features is the development of a strong anisotropy in the form of geostrophic vortices which are invariant along the rotation axis and correspond to a balance between pressure gradients and the Coriolis force \cite{godeferd_structure_2015}.
It is often thought that, in the asymptotic regime of rapid rotation, geostrophic modes should be the dominant component of rotating flows because they are unable to generate significant three-dimensional perturbations \cite{vanneste_balance_2013,gallet_exact_2015}.
Yet, this picture has recently been questioned by a number of experimental and numerical works which have proved the existence of purely three-dimensional rotating turbulence.
Transition between two- and three-dimensional rotating turbulence has been found in direct numerical simulation by changing the amplitude of the forcing \cite{yokoyama_hysteretic_2017,van_kan_critical_2020} or by changing the aspect ratio of the simulated domain \cite{van_kan_critical_2020}.
Two experimental studies \cite{le_reun_experimental_2019,brunet_shortcut_2020} have focused on the turbulence resulting from forcing inertial waves, which are three-dimensional waves that propagate under the restoring action of the Coriolis force. 
They have shown that exciting waves at a particular frequency excites a series of new inertial waves via non-linear, resonant, triadic interactions, provided that the forcing amplitude is sufficiently small; at larger forcing, a two-dimensional geostrophic state is recovered. 
Lastly, rotating Rayleigh-B\'enard convection has also been shown to sustain two- and three-dimensional turbulent states \cite{favier_subcritical_2019}.
Such transitions are important to characterise on a fundamental level as they change the dissipative properties of the flow. 
The dissipation anomaly and the forward energy cascade that are typical of three-dimensional turbulent flows disappear when two-dimensional motions dominate \cite{campagne_direct_2014}; instead, energy accumulates in large scales under the action of an inverse cascade \cite{alexakis_cascades_2018}. 

As shown in numerical simulations performed in less dissipative regimes than in experiments, injecting energy through only inertial waves drives inertial wave turbulence \cite{yarom_experimental_2014,le_reun_inertial_2017}, a state where many waves interact resonantly and create a direct energy cascade towards smaller scales \cite{galtier_weak_2003,nazarenko_wave_2011}. 
Wave turbulence has been characterised experimentally or numerically in various media supporting linear dispersive waves. 
Examples include gravity-capillary waves at the surface of water \cite{falcon_observation_2007,deike_direct_2014,aubourg_nonlocal_2015}, flexural waves in plates \cite{during_weak_2006,
miquel_nonstationary_2011}, Alfv\'en waves in plasmas \cite{galtier_spectral_2005} or internal gravity waves in stratified fluids \cite{brouzet_energy_2016,le_reun_parametric_2018}.
Unlike classical homogeneous, isotropic turbulence, the spatial spectrum of wave turbulence (the \ZK spectrum) can be predicted from asymptotic analysis of the Navier-Stokes equations.
In the case of inertial waves, the \ZK spectrum has been derived by Galtier \cite{galtier_weak_2003} assuming that non-linear interactions are local and lead to strong anisotropy by enhancing small, but non-zero, vertical wavenumbers.
Yet, such a spectrum has never been observed in direct numerical simulations or experiments. 
It was indirectly observed in \cite{bellet_wave_2006} with an asymptotic model computing the turbulent energy transfers, but the relevance of the \ZK spectrum for experiments and direct numerical simulations remains to be evaluated. 
In this article, we address this issue with long simulations performed at low forcing amplitude and dissipation rate which thus draw closer to the adequate asymptotic limit for wave turbulence. 
The turbulent flow is forced with either isotropic random excitation or  with a linear, resonant excitation of waves only.
To retrieve the purely three-dimensional states reported in \cite{le_reun_experimental_2019,
van_kan_critical_2020,
brunet_shortcut_2020}, a friction is applied specifically to geostrophic modes as done numerically in \cite{le_reun_inertial_2017} and then realised experimentally in \cite{brunet_shortcut_2020}. 
We show that our direct numerical simulations are in good agreement with the \ZK spectrum of wave turbulence predicted in \cite{galtier_weak_2003} and with asymptotic models \cite{bellet_wave_2006}. 
Lastly, we show by releasing the friction that rotating turbulence has two equilibrium states for the same control parameters, geostrophic turbulence and wave turbulence.

\section{Methods}

\begin{figure*}
\centering
\includegraphics[width=\linewidth]{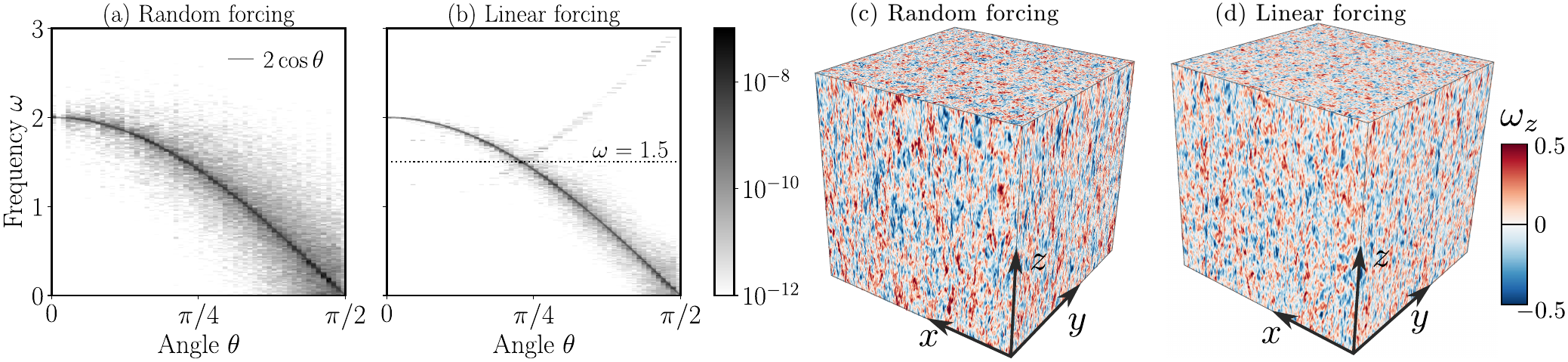}
\caption{\textbf{a and b:} Spatio-temporal spectra $\mathcal{E}(\theta,\omega)$ for the two most extreme simulations ($E = 10^{-7.5}$, resolution $512^3$) using random (a) and linear (b) forcing.
The continuous line in panel a shows the dispersion relation of inertial wave \ref{eq:dispersion_relation}.
The value and the location of the forcing frequency is specified in panel b.
In panel b, the secondary locations of energy mirroring the dispersion relation are bound waves due to non-linear, non-resonant interaction between inertial waves and the forcing flow.
\textbf{c and d:} snapshots of the vertical vorticity $\omega_z$ taken from simulations using random (c) and linear (d) forcing. 
 }
\label{fig:spatio_temporal_spectra}
\end{figure*}

%\subsection{Modeling equations}
%
Rotating turbulence is investigated through direct numerical simulations in a cubic box of size $L$ with periodic boundary conditions.
Time is normalised by the global rotation rate $\Omega$ and distance by the size of the box. 
The dimensionless, rotating Navier-Stokes equations which govern the flow write 
\begin{equation}
\label{eq:NS_perturbation}
\begin{array}{rl}
\partial_t \bu +  \bu \cdot \bnabla \bu + 2 \mbf{e}_z \times \bu  &= - \bnabla \Pi + E \bnabla^2 \bu + \mathcal{F}\left[\bu \right] \\
\bnabla \cdot \bu &= 0 ~.
\end{array}
\end{equation}
where we have introduced the modified pressure $\Pi$ and the Ekman number $E = \nu/(L^2 \Omega)$, $\nu$ being the viscosity, which quantifies the effects of viscous drag over the Coriolis force. 
Turbulence is forced through the term $\mathcal{F}[\bu]$ which may be a function of the flow $\bu$ itself. 
The flow is simulated by decomposing $\left\lbrace \bu, \Pi \right\rbrace$ into plane waves $\left\lbrace \hat{\bu}_k, \hat{p}_k \right\rbrace \exp i (\bk \cdot \mbf{x})$, which is allowed by the periodic boundary conditions.
The equations \ref{eq:NS_perturbation} are solved using the pseudo-spectral code \textsc{Snoopy} \cite{lesur_relevance_2005} where the non-linear term is computed in the real space with a standard $2/3$ de-aliasing process. 
Two different types of forcing functions $\mathcal{F}[\bu]$ are considered. 
In a first set of simulations, $\mathcal{F}$ is a noise that is $\delta$-correlated in time and applied to the modes $ \bk$ located in a spherical shell with inner and outer radii $k_f = 2 \pi \times 10$ and $k_f + 2 \pi$, respectively.
\revision{Such a value for $k_f$ imposes a scale separation between the forcing and the size of the simulated domain, thus limiting finite-size effects and favouring the emergence of wave turbulence \cite{denissenko_gravity_2007}.}
In a second set of simulations, we use a linear, resonant forcing that excites only inertial waves through a parametric resonance.
Linear and time-coherent forcing of turbulence has recently proved a useful tool to better control the energy injection process \cite{le_reun_inertial_2017,supekar_linearly_2020,
qin_transition_2020}.
In the present case, the linear forcing is
\begin{equation}
\label{eq:define_elliptical_forcing}
\mathcal{F}[\bu] = - Ro_i\left(\bU_b \cdot \bnabla \bu - \bu \cdot \bnabla \bU_b\right)
\end{equation}
\begin{equation}
\label{eq:define_Ub}
\mbox{with:}~~\mbf{U}_b = -\bnabla \left[ \frac{1}{2} \sin(2\gamma t) (x^2 - y^2) + \cos(2 \gamma t) xy \right]
\end{equation}
and $Ro_i$ a dimensionless measure of the forcing flow amplitude.
\revision{It may be seen as an input Rossby number since, at steady state, the balance $\mathcal{F}[\bu] \sim \bu \cdot \bnabla \bu$ imposes $\vert \bu\vert  \propto Ro_i$.}
$\bU_b$ is a rotating strain flow well-known in geo- and astrophysics as the response of a planetary or stellar fluid interior to tidal excitation \cite{kerswell_elliptical_2002,le_bars_flows_2015}.
As shown in many previous studies, this flow is unstable and drives parametric resonance of inertial waves at frequency $\omega \simeq \gamma$ which then grow exponentially, a mechanism called the elliptical instability \cite{kerswell_elliptical_2002}.
The inviscid growth rate of waves is maximum for modes such that $k_y \simeq \pm k_x$ and is independent of $\vert \bk\vert$ \cite{kerswell_elliptical_2002}.
The mode selection on the discretised Cartesian grid is achieved through a competition between resonance accuracy ($\vert \omega - \gamma \vert = O(Ro_i)$) and viscous damping.
In the non-linear saturation of the instability, such a forcing $\mathcal{F}[\bu]$ thus allows injecting energy via a small number of inertial waves only. 
\revision{
Although the forcing flow $\bU_b$ is not space-periodic, its linearity allows to simulate its effect via the shearing box method \cite{rogallo_numerical_1981}: 
the effects of the term $\bU_b \cdot \bnabla \bu$ are accounted for by imposing $O(Ro_i)$ periodic oscillations of the wavevectors around their mean \cite{barker_non-linear_2013}. 
}

As we are interested in the three-dimensional states of rotating turbulence, we apply a friction specific to geostrophic modes to force them to remain of small amplitude as done previously in numerical \cite{le_reun_inertial_2017} and experimental \cite{brunet_shortcut_2020} studies.
This is done in the spectral space by adding a strong damping term $-\hat{\bu}_k$ to the evolution equation of the modes with wave vectors $\bk$ such that $\kpar = \bk \cdot \mbf{e}_z = 0$.
To investigate inertial wave turbulence in the asymptotic regime of rapid rotation, we proceed to simulations where the Ekman number is varied from $10^{-6.5}$ to $10^{-7.5}$, with resolution ranging from $256^3$ to $512^3$ modes. 
In the case of the linear forcing (see equation \ref{eq:define_elliptical_forcing}), the input Rossby number is set to $Ro_i = 7.5 \times 10^{-3}$ and the frequency to $\gamma = 1.5$. 
For this set of parameters, the unstable wavenumbers excited by the linear forcing are close to $k_f$, the forcing wavenumber in the random forcing simulations.
\revision{In addition, the values for $E$ and $Ro_i$ ensure that the viscous damping $\sim k_f^2 E$ is smaller than the growth rate of the instability $\sim Ro_i$.}
When random forcing is used, the noise amplitude is tuned to ensure that the root mean square (rms) amplitude of the steady state flow matches the one obtained with the linear forcing. 
All these simulations reach a statistically steady state that is simulated over at least 40 non-linear timescales, \revision{the latter being given by  $Ro_i^{-1}$ since $\vert \bu \vert \propto Ro_i$}.
\revision{Such a long time is necessary to reach a good convergence of the statistical properties of inertial wave turbulence and is the main limiting factor numerically.
}

Lastly, we use a reference simulation with linear forcing where there is no geostrophic friction and for which geostrophic modes dominate the dynamics.
The Ekman number is set to $10^{-7.5}$ and the input Rossby number is $Ro_i = 3 \times 10^{-3}$. 
Such a low value for $Ro_i$ allows to have matching rms velocity with the wave-dominated simulations.
As noted in references \cite{barker_non-linear_2013} and \cite{le_reun_inertial_2017}, the turbulent flow resulting from the linear excitation consists in cycles of wave instability followed by slow viscous decay where the large-scale geostrophic component dominates and the action of the forcing is suppressed by strong detuning. 
Statistical properties of the flow are computed during a decay phase that lasts 30 non-linear timescales and over which geostrophic modes lose virtually no energy. 

\section{Wave turbulence and anisotropic spectra}

\begin{figure*}
\centering
\includegraphics[height=0.215\linewidth]{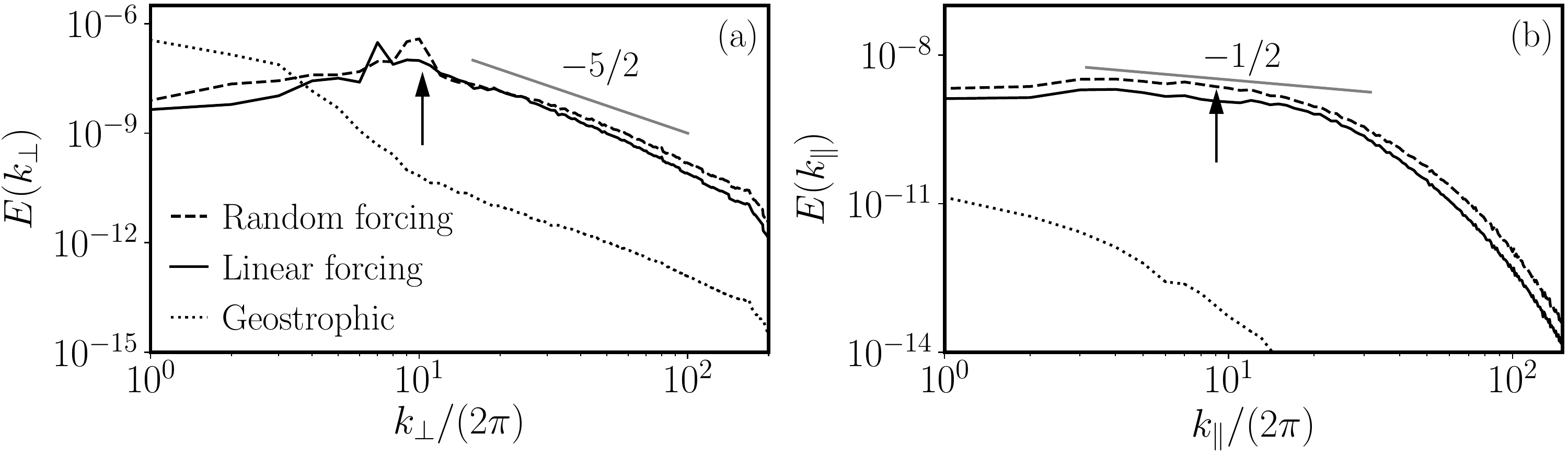}
\includegraphics[height=0.215\linewidth]{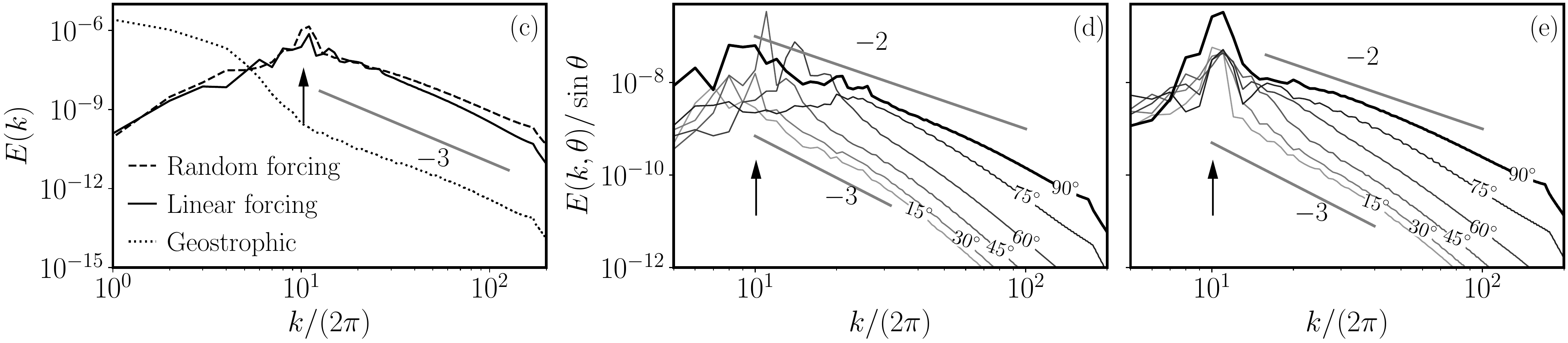}
\caption{
\textbf{a and b:} Anisotropic spectra $E(\kper)$ and $E(\kpar)$ as defined in equations \ref{eq:def_Ekper} and \ref{eq:def_Ekpar}, the expected power laws are given with grey plain segments. 
\textbf{c:} Isotropic spectra compared to the $k^{-3}$ law expected from \cite{bellet_wave_2006}.
\textbf{d and e:} Angular spectra $E(k,\theta)$ from linear (d) and random (e) forcing. 
Each curve is labelled with the maximum angle $\theta$, and energy is averaged over a range of $15^\circ$. 
The spectra are compensated by the relative number of modes, $ \sin \theta$. 
The limiting power laws $k^{-2}$ and $k^{-3}$ for $E(k,\theta)$ are indicated on both panels. 
On all panels, the vertical arrow materialises the injection scale.
}
\label{fig:Galtier_spectra}
\end{figure*}

Typical snapshots of the three-dimensional flows driven by linear and random forcings and with geostrophic friction are shown in \ref{fig:spatio_temporal_spectra}.
There is a clear production of smaller scales from a larger-scale forcing.
It is associated to a dissipation anomaly: with linear forcing, as the Ekman number is decreased from $10^{-6.5}$ to $10^{-7.5}$, the dissipation (defined as the volume average of $-E \vert \bnabla \bu \vert^2$) only decreases from $2 \times 10^{-9}$ to $1.3 \times 10^{-9}$.
As the typical rms velocity is $u_{\mathrm{rms}} \sim 10^{-3}$, the dissipation is of the same order of magnitude as the dimensionless inertial estimate $u_{\mathrm{rms}}^3$.
The small-scale, three-dimensional turbulent flow is a superposition of inertial waves. 
Their presence is assessed by a projection of the kinetic energy in the sub-space of the dispersion relation \revision{ \cite{yarom_experimental_2014,clark_di_leoni_quantification_2014,clark_di_leoni_spatio-temporal_2015,aubourg_nonlocal_2015,
brouzet_energy_2016,le_reun_inertial_2017,lam_partitioning_2020}}.
In the case of inertial waves, the dispersion relation links the frequency $\omega$ to the angle $\theta$ between the wavevector $\bk$ and the rotation axis $\mbf{e}_z$ as follows,
\begin{equation}
\label{eq:dispersion_relation}
\omega^2 = 4 \cos^2 \theta = 4 \kpar^2/ k^2.
\end{equation} 
In the spectral space, the amplitudes $\hat{\bu}_k (t)$ are summed over the wavevectors with the same $\theta$ which gives a quantity $\hat{\bu}(\theta,t)$. 
A temporal Fourier transform is then applied in the steady state to obtain $\bu (\theta,\omega)$; the desired energy representation is then $\mathcal{E}(\theta,\omega) = \vert \bu(\theta,\omega)\vert^2$. 
The spatio-temporal spectra $\mathcal{E}(\theta,\omega)$ are shown in \ref{fig:spatio_temporal_spectra}.
The kinetic energy is clearly focused along the dispersion relation of inertial waves for both forcings \revision{with $95$ \% of the energy being contained within a frequency distance of $4.5 Ro_i$ (random) or $2.5Ro_i$ (linear) to the exact dispersion relation.}
When the flow is forced through the linear instability (\ref{fig:spatio_temporal_spectra}b), energy is injected only at the intersection between the dispersion relation and the resonance frequency $\gamma$.
Other structures lying along the dispersion relation must be excited through triadic resonances, a non-linear process by which three waves exchange energy over long time scales.
They occur provided that their wavenumbers ($\bk_1$, $\bk_2$, $\bk_3$) satisfy $\bk_1 + \bk_2 + \bk_3 = \mbf{0}$ and that their frequencies $\omega_i$ satisfy \revision{$\omega_1 + \omega_2 + \omega_3 = O(Ro_i)$ \cite{vanneste_wave_2005}}. %\cite{bordes_experimental_2012-1} 
When random forcing (\ref{fig:spatio_temporal_spectra}a) is used, waves may be directly excited by the forcing and it becomes more difficult to tell apart the role of non-linearity in filling the dispersion relation.
\revision{Moreover, since energy is randomly injected into modes rather than at their eigenfrequencies, the focusing along the dispersion relation is broader with stochastic forcing.
}
The statistical properties of the three-dimensional turbulence examined here is tested against theories and asymptotic models of inertial wave turbulence. 
The first prediction that we consider here is the \ZK spectrum of kinetic energy derived for the weak wave turbulence theory
\cite{zakharov_kolmogorov_2012,nazarenko_wave_2011}.
It was derived by Galtier \cite{galtier_weak_2003} who found it to be strongly anisotropic and to depend on both the vertical and the horizontal components of the wavevector, $k_{\perp} = k \sin \theta$ and $k_{\parallel} = k_z = k \cos \theta$.
The kinetic energy spectrum could be derived in the limit $\kpar \ll \kper$ where it takes the following form,
\begin{equation}
\label{eq:Galtier_spectrum}
E(\kper,\kpar) \propto \kpar^{-1/2} \kper^{-5/2}~.
\end{equation}
The anisotropic density of kinetic energy is obtained from our simulations by summing the $\vert \hat{\bu}_k\vert^2$ at constant $k$ and $\theta$ within a small tolerance $\Delta k = 2\pi$ and $\Delta \theta = \pi/120$, respectively. 
To provide a quantitative comparison with \ref{eq:Galtier_spectrum}, we compute two one-dimensional, integrated spectra, $E(\kper)$ and $E(\kpar)$ defined as
\begin{align}
E(\kper) &= \int_{\kpar = 0}^{k_f} E(\kper,\kpar) \mathrm{d} \kpar \label{eq:def_Ekper} \\
E(\kpar) &= \int_{\kper = 5 k_f}^{\infty} E(\kper,\kpar) \mathrm{d} \kper.\label{eq:def_Ekpar}
\end{align}
The bounds of the integrals \ref{eq:def_Ekper} and \ref{eq:def_Ekpar} are chosen to be consistent with the limit $\kpar \ll \kper$.
In practice, changing the upper bound in \ref{eq:def_Ekper} does not affect the spectrum because energy is mostly located at small $\kpar$. 
However, decreasing the lower bound of the integral \ref{eq:def_Ekpar} too close to $k_f$ strongly steepens the spectrum.
From the prediction \ref{eq:Galtier_spectrum}, $E(\kper) \propto \kper^{-5/2}$  is expected at large wavenumbers whereas $E(\kpar) \propto \kpar^{-1/2}$ is expected at small wavenumbers. 
The integrated spectra are shown in \ref{fig:Galtier_spectra}a and \ref{fig:Galtier_spectra}b.
Both the linear and random forcings produce a $\kper^{-5/2}$ power law. 
The $\kpar^{-1/2}$ is less convincing, and it seems that drastically larger $k_f$ would be required to observe its full development.
In the simulation without friction and with dominant geostrophic modes, the energy $E(\kper)$ has accumulated in larger scales under the action of the inverse cascade.
It still shows a $-5/2$ power law even though the energy is much smaller due the disruption of the forcing by geostrophic modes \cite{barker_non-linear_2013,le_reun_inertial_2017}.
However, the spectrum $E(\kpar)$ falls more quickly than in the wave turbulence simulations: energy has condensed in the geostrophic plane and wave turbulence is suppressed. 
The $\kper^{-5/2}$ may only be a reminiscence from the short-lived resonance that precedes the geostrophic-dominated decay and that has shortly excited smaller-scale waves. 
It is also plausible that inertial wave turbulence can still develop in a certain form even when the geostrophic flow dominates, a possibility that is beyond the scope of the present study.
\revision{
For instance, spectra close to the prediction of Galtier \ref{eq:Galtier_spectrum} have recently been reported in simulations of randomly forced rotating turbulence where wave and vortices co-exist \cite{yokoyama_energy-flux_2020}.
}

Our direct numerical simulations are also tested against the asymptotic model of Bellet \textit{et al} \cite{bellet_wave_2006}. 
These authors investigated inertial wave turbulence using 
numerical computation of resonant energy transfers between waves with an Eddy-Damped Quasi-Normal Model (EDQNM) type of closure. 
\revision{Their simulations were initiated by an isotropic homogeneous turbulent state that relaxes under the influence of rotation.
Over the slow decay phase, the flow becomes an inertial wave turbulence whose statistical properties were investigated.}
The isotropic spectrum was found to follow a $k^{-3}$ power law, which we also report in our simulations with linear forcing (see figure \ref{fig:Galtier_spectra}c). 
In the case of random forcing, the isotropic spectrum is closer to $k^{-2}$  just above the forcing wavenumber. 
Furthermore, Bellet et al \cite{bellet_wave_2006} computed the kinetic energy spectrum as a function of $k$ and $\theta$, $E(k,\theta)$, which is a polar coordinate formulation of the spectrum $ E(\kper,\kpar)$ such that $E(k,\theta) \sim k E(\kper(k,\theta),\kpar(k,\theta))$.
They reported that $E(k,\theta \rightarrow 0 )$ decays faster than $k^{-3}$ and $E(k,\theta \rightarrow \pi/2 ) \propto k^{-2}$.
In the present DNS, the spectra $E(k,\theta)$ are computed by summing the kinetic energy at constant $k$ and constant $\theta$ within a tolerance of $15^\circ$. 
They are then compensated by the relative number of modes in each angular area, $\sin \theta$. 
With linear forcing (\ref{fig:Galtier_spectra}d), a $k^{-2}$ law is observed for the largest angles close to $\pi/2$, and steeper than $k^{-3}$ laws are reported for the smallest angles. 
As noted in \cite{bellet_wave_2006}, the $k^{-2}$ spectrum is consistent with the prediction \ref{eq:Galtier_spectrum}.
Considering $\theta$ constant and close to $\pi/2$, $\kper \sim k$ and $\cos \theta = \kpar/\kper$ is constant.
In this limit, the theoretical prediction \ref{eq:Galtier_spectrum} becomes $E(k,\theta \rightarrow \pi/2)\sim k  k^{-5/2}  k^{-1/2}= k^{-2}$.
The spectral behaviour of the flow created by random forcing is overall the same, but the largest $\theta$ spectrum tends to be shallower towards $k_f$.
It is worth noting in \ref{fig:Galtier_spectra}e that the modes with $\theta \rightarrow \pi/2$ absorb relatively more energy from the isotropic random forcing. 
Under local transfers, this anomalous amount of energy remains in the same angular area which also receives energy from larger $\theta$ modes \revision{thus making them relatively more energetic than in the linear forcing case.}
\revision{The inertial wave turbulence is thus more anisotropic with stochastic forcing.}
\revision{Lastly, a $k^{-2}$  power law for the isotropic spectrum has been predicted in \cite{zhou_phenomenological_1995} and \cite{galtier_weak_2003}, but under the assumption of isotropy of the flow.
The latter hypothesis is clearly incompatible with the spectra of \ref{fig:Galtier_spectra}e. and cannot explain the shallower spectra found with random forcing.}
To conclude, our DNS of geostrophic-free rotating turbulence produce inertial wave turbulence whose statistical properties are in good agreement with the theoretical predictions drawn from the wave turbulence theory, in particular the \ZK spectrum. 
The agreement tends to be better when linear forcing of waves is used rather than random forcing. 
The simulation with linear forcing and free geostrophic modes shows that energy is mostly contained in large scales. 
Reminiscence of wave turbulence is observed in the $E(\kper)$ spectrum, but energy has collapsed in geostrophic modes, which makes this state fundamentally different from wave turbulence. 
Further discussion of these differences can be found in reference \cite{le_reun_inertial_2017}.
\revision{Despite being promising, it is important to note that these simulations are performed in a regime that remains influenced by dissipation, a caveat that is common to many wave turbulence studies, even in two-dimensional set-ups \cite{mordant_fourier_2010,deike_direct_2014}.
There is, indeed, less than an order of magnitude of separation between the non-linear timescale based on the rms velocity ($u_{\mathrm{rms}}^{-1} \sim 10^3$) and the viscous dissipation timescale ($(k_f^2 E)^{-1} \sim 8 \times 10^3$). 
However, achieving larger timescale separation remains difficult with the currently available computing power given the long time needed for statistics to converge. 
}

\section{Spectral energy transfers}

\begin{figure}
\centering
\includegraphics[width=\linewidth]{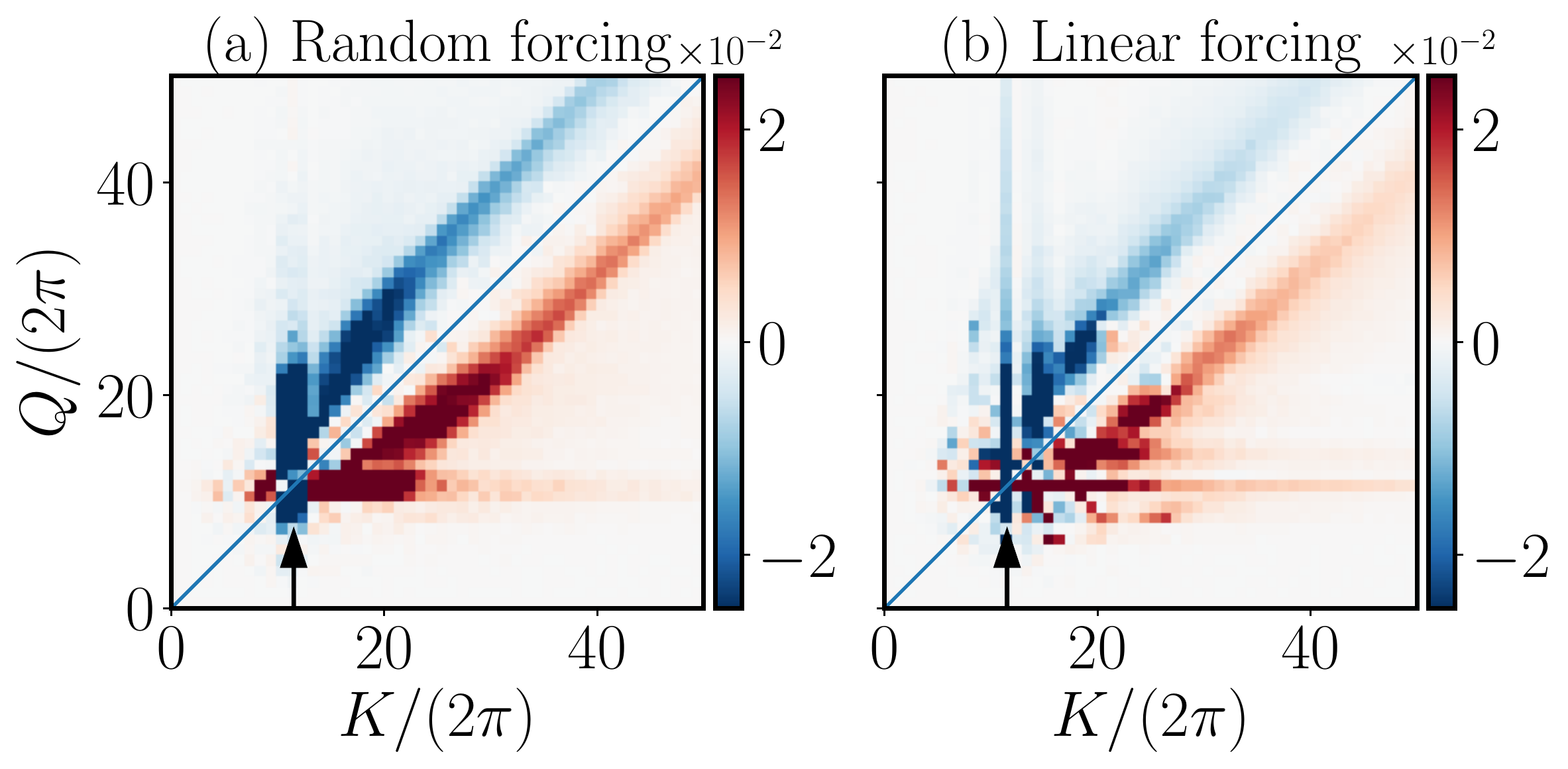}
\caption{Signed heat map of the kinetic energy transfer from spectral shell $Q$ to shell $K$. The central line materialises the bisector $Q=K$ and the vertical arrow marks the injection scale. The transfers are normalised by $u_{rms}^3$. }
\label{fig:transfer_maps}
\end{figure}

The derivation of the \ZK spectrum (as in equation \ref{eq:Galtier_spectrum}) is based on the assumption of locality of wave interactions for which energy is transferred between wavevectors of similar modulus.
%
%The observation of the wave turbulence spectrum is thus a signature of these interactions in the present simulations.  
%
As noted in \cite{waleffe_inertial_1993,cambon_energy_1997,galtier_weak_2003}, local
interactions are necessarily anisotropic and directed towards modes with smaller $\kpar$ and smaller frequencies. 
The observation of the \ZK spectrum is a signature of local interactions, but we also find in \ref{fig:spatio_temporal_spectra}b that energy is also transferred to higher frequencies, thus suggesting the coexistence of non-local transfers.
Local and non-local transfers can be identified through the approach of Alexakis et al \cite{alexakis_shell_2005} by computing the shell-to-shell transfer function $\TKQ$ defined as
\begin{equation}
\TKQ(K,Q) = \int_{\mathcal{V}} \bu_K \cdot \left( \bu \cdot \bnabla \bu_Q  \right) \mathrm{d}^3 \mbf{x}
\end{equation}
where the integral is performed over the volume of the cube and the $\bu_P$ are defined as
\begin{equation}
\bu_P = \sum_{P\leq \vert \bk \vert< P+2 \pi} \hat{\bu}_k e^{i \bk \cdot \mbf{x}}~.
\end{equation}
The transfer function $\TKQ(K,Q)$ then quantifies the transfer from the spectral shell with radius $Q$ to the shell with radius $K$, and is an anti-symmetric quantity.  
The transfer map $\TKQ(K,Q)$ is shown in \ref{fig:transfer_maps} in the case of random and linear forcing.
Local transfers occur close to the bisector $Q = K$ and are clearly identifiable in both panels of \ref{fig:transfer_maps}. 
Since the transfer is overall positive below the bisector ($K\geq Q$) the cascade is directed towards smaller scales, as expected.
Perfectly local transfers, as in classical turbulence, would peak very close to the bisector \cite{alexakis_shell_2005}. 
Here, we observe a systematic gap between the wavenumbers $K$ and $Q$ of modes undergoing local transfer. 
The origin of this gap and its possible relation with the complexity of resonant surfaces remain to be understood.
In addition, a strongly non-local cascade is observed in the transfer maps of \ref{fig:transfer_maps} and is characterised by lines of significant transfer at constant $K$ or $Q$.
They occur mostly at the injection scale and are more important in the linear forcing case where energy is injected through waves only. 
If they were to persist in the asymptotic limit of vanishing Rossby numbers, the systematic wavenumber gap and the non-local interactions could be important indications to extend the theory of inertial wave turbulence.
Following the discussion in \cite[p.3]{galtier_weak_2003}, these results suggest that a cascade towards smaller scale but with $\kpar \sim \kper$ can exist in inertial wave turbulence. 
Although it does not affect the modes with $\kpar \ll \kper$ (hence the good agreement of our data with the prediction \ref{eq:Galtier_spectrum}), such a cascade will possibly be important to make analytical predictions outside this limit.

\section{Bi-stability of rotating turbulence}

\begin{figure}
\centering
\includegraphics[width=\linewidth]{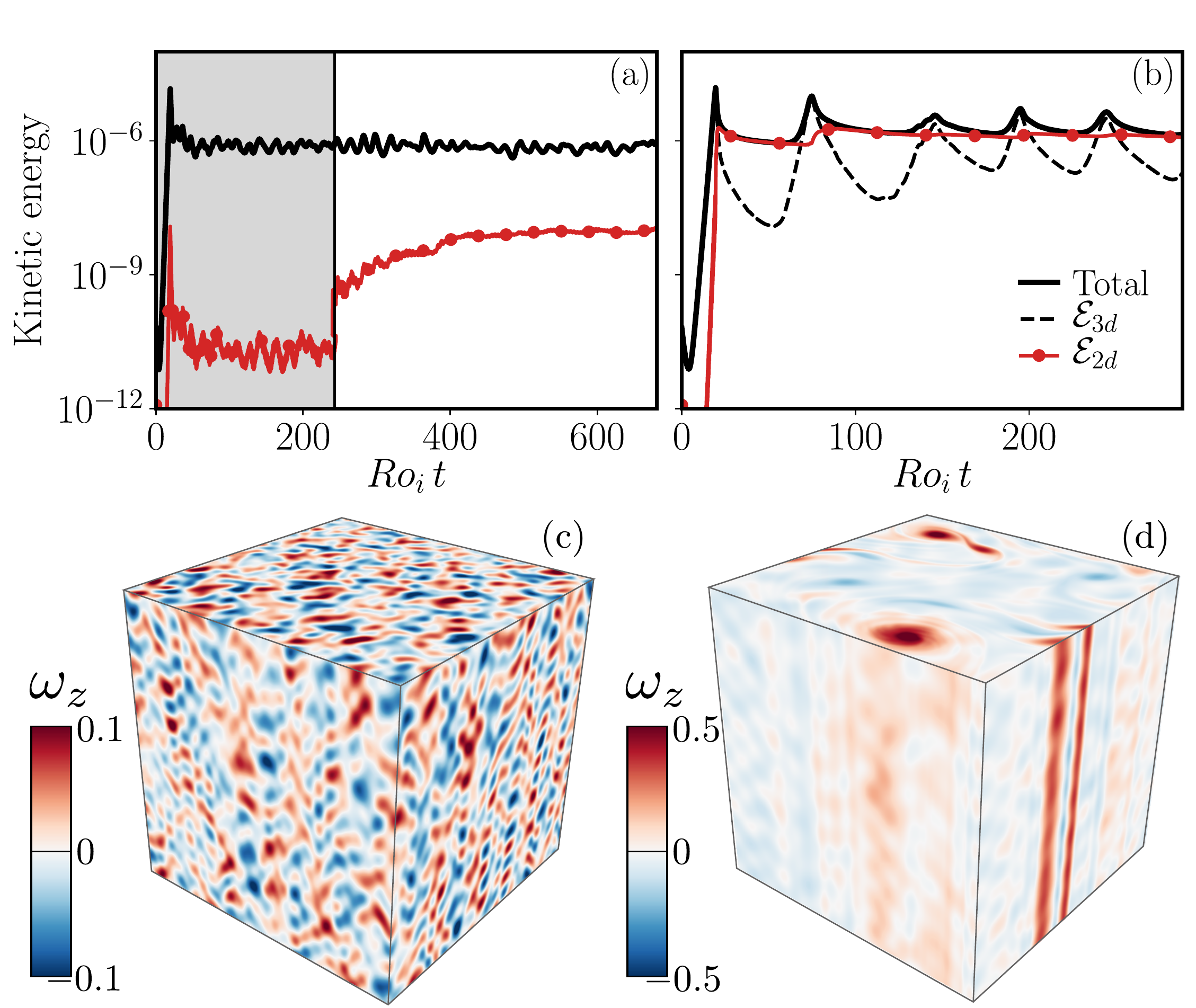}
\caption{(a and b) Times series of the kinetic energy for two simulations using linear forcing both carried out at $Ro_i = 7.5 \times 10^{-3}$and $E = 10^{-6.5}$. 
In the first simulation (a), geostrophic friction is first applied (grey area) and then released, whereas in the second (b), no friction is applied. 
The corresponding snapshots of the vertical vorticity are shown in panels c and d and taken at times $Ro_i t = 500$ and $Ro_i t = 90$ respectively.
%
%\revision{Time is normalised by the non-linear timescale $Ro_i^{-1}$.}
%
}
\label{fig:bistability}
\end{figure}

%In this last section, 
Lastly, we explore the stability of the inertial wave turbulence that has been created in our simulations with an artificial damping by releasing the latter after a certain time. 
Such a process is carried out in the case of the linear forcing which does not directly excite modes with small $\kpar$ that are near the geostrophic plane. 
Simulations are carried out at the same input Rossby number ($Ro_i = 7.5 \times 10^{-3}$) but at larger Ekman number ($E = 10^{-6.5}$) and lower resolution ($N=256$), which allows solving the long-time dynamics (typically up to 500 non-linear timescales).
Geostrophic friction is first applied to create an inertial wave turbulence that has reached a statistically steady state.
The friction is then released and the system evolves towards a new equilibrium state where the geostrophic modes are unconstrained.
The outcome of such a process is shown in \ref{fig:bistability}a and c. 
The respective time evolution of waves and geostrophic modes is tracked by decomposing the kinetic energy into a geostrophic (modes with $\kpar = 0$, $\Edd$) and a wave (modes with $\kpar \neq 0$, $\Eddd$) contributions.  
Once the friction is released, after a quick adjustment, the geostrophic kinetic energy $\Edd$ slowly increases to reach a new steady state where it is at least two orders of magnitude below the wave energy $\Eddd$. 
Inertial wave turbulence is thus stable for these values of the Ekman and the Rossby numbers, at least for the duration of the present simulation. 
Interestingly, when a simulation is carried out with the same parameters but without geostrophic friction at all, a non-linear state with strong geostrophic component is recovered (see \ref{fig:bistability}b and d), a state well documented in references \cite{barker_non-linear_2013} and \cite{le_reun_inertial_2017}.
For $Ro_i = 7.5 \times 10^{-3}$ and $E = 10^{-6.5}$, rotating turbulence excited by the linear forcing is thus bi-stable.
Bi-stability is lost when the Ekman number is reduced to $10^{-7}$: as friction is released, the geostrophic component becomes unstable, grow exponentially and ends up dominating the flow.
\revision{It is also lost when random forcing is used since energy is then directly injected into geostrophic modes.
Without friction, because of the inverse cascade, energy accumulates in large-scale vortical flows that become predominant.}
\revision{With linear forcing,} viscosity is a possible candidate for the stabilisation of geostrophic modes at $E= 10^{-6.5}$ and $Ro_i = 7.5 \times 10^{-3}$. 
A crude estimate of the inviscid growth rate of the geostrophic instability can be obtained from the threshold of the destabilisation. 
The geostrophic instability vanishes when the viscous damping $k^2 E$ balances the inviscid growth rate.
Taking $k \sim k_f$, we obtain $E k^2 \sim  10^{-3}$ as a crude estimate of the inviscid growth rate.
It may be compared to the growth rate of inviscid geostrophic instabilities, namely the quartetic \cite{kerswell_secondary_1999,brunet_shortcut_2020} and the near-resonant \cite{le_reun_near-resonant_2020} instabilities. 
They are both $\sim (k_f Ro)^2$ (to an order one prefactor) where $Ro$ is the rms amplitude of the waves.
Taking $k_f = 2 \pi \times 10$ and $Ro \sim 10^{-3}$ (see \ref{fig:bistability}), we find that $(kRo)^2 \sim 3 \times 10^{-3}$, which is of the same order of magnitude as the estimated inviscid growth rate. 
The geostrophic instability that we observe here is thus consistent with the two mechanisms that have been recently described \cite{brunet_shortcut_2020,le_reun_near-resonant_2020}, although it is not possible to distinguish them with the available data.
Yet, the origin and the domain of existence of bi-stability of rotating flows excited by inertial waves remain to be described.

\section{Conclusion}

We have carried out numerical simulations of rotating turbulence in which the geostrophic flow is forced to be subdominant in a regime of low forcing amplitude and dissipation that had not been explored yet.
Such a turbulent state bears all the characteristics of weak inertial wave turbulence as the flow is comprised only of waves in weak non-linear interaction.
We have found that it follows the prediction drawn from the wave turbulence theory, namely the \ZK spectrum derived by Galtier \cite{galtier_weak_2003}.
It also agrees well with the asymptotic EDQNM models of inertial wave turbulence where only resonant interactions are computed \cite{bellet_wave_2006}. 
These predictions tend to be more accurate when linear forcing of waves is used instead of isotropic random forcing; injecting energy into waves only is thus more adequate to observe inertial wave turbulence.
Our study thus provides a \revision{promising} example of the success of the wave turbulence theory which provides adequate prediction even in the present case that is strongly anisotropic and three-dimensional. 
\revision{
Nevertheless, despite our set-up being optimised for the observation of wave turbulence, the states observed here remains influenced  by dissipation due to the inherent limitations of the numerics. 
A fully-developed inertial range remains to be observed but seems presently out of the reach of systematic exploration. 
}
\revision{Lastly, the use of an artificial friction has allowed us to show that rotating turbulence can be bi-stable when it is forced via inertial waves.}
For certain values of $E$ and $Ro_i$, wave turbulence remains stable after the release of friction whereas a geostrophic-dominated state is retrieved when no friction is used from the start.
It thus brings a new proof that geostrophic states are not universal to rotating turbulence.
% and it is a further argument in favour of the use of artificial friction to probe new turbulent states. 
%
%In addition to making possible the study of a three-dimensional turbulent state that is relevant to experiments and geophysics \cite{le_reun_experimental_2019,le_reun_inertial_2017}, it has allowed us to show that such a state can be stable in the same condition as the geostrophic turbulence. 
%
%Yet, the boundaries of the respective stability areas of geostrophic and wave turbulence in terms of the control parameters (forcing and dissipation) remains to be fully characterised, a task that will require a large numerical and theoretical effort. 
%
\revision{It will have important implications in geo- and astrophysics.
For instance, when turbulence is excited by tidal interaction in stellar and planetary interiors \cite{le_bars_flows_2015}, transitions between two- and three-dimensional turbulence will strongly influence dissipation---and thus orbital evolution---as well as magnetic field generation. 
}

%\vspace{-0.5cm}

\acknowledgments
The authors acknowledge funding by the European Research Council under the European Union's Horizon 2020 research and innovation program through Grant No. 681835-FLUDYCO-ERC-2015-CoG. TLR is supported by the Royal Society through a Newton International Fellowship (Grant reference NIF\textbackslash R1\textbackslash 192181). Numerical simulations were performed using HPC resources from GENCI-IDRIS (Grant 2020-10080407543).
%\vspace{-0.5cm}

%\bibliographystyle{eplbib}
%\bibliography{biblio}

\end{document}